\documentstyle[floats,aps,prb]{revtex}

\begin{document}

\title{Continuum Elastic Theory of Adsorbate Vibrational Relaxation}
\author{Steven P. Lewis,$^a$ M. V. Pykhtin,$^b$ E. J. Mele,$^b$ 
        and Andrew M. Rappe$^a$}
\address{Department of Chemistry,$^a$ Department of Physics,$^b$ and \\
         Laboratory for Research on the Structure of Matter,$^{a,b}$ \\ 
         University of Pennsylvania, Philadelphia, PA 19104.}
\date{\today}
\maketitle

\begin{abstract} 
An analytical theory is presented for the damping of low-frequency
adsorbate vibrations via resonant coupling to the substrate phonons.
The system is treated classically, with the substrate modeled as a
semi-infinite elastic continuum and the adsorbate overlayer modeled
as an array of point masses connected to the surface by harmonic springs.  
The theory provides a simple expression for the relaxation rate in terms
of fundamental parameters of the system: 
$\gamma = m\bar{\omega}_0^2/A_c \rho c_T$, where $m$ is the adsorbate mass,
$\bar{\omega}_0$ is the measured frequency, $A_c$ is the overlayer unit-cell 
area, and $\rho$ and $c_T$ are the substrate mass density and transverse 
speed of sound, respectively.  This expression is strongly coverage dependent, 
and predicts relaxation rates in excellent quantitative agreement with 
available experiments.  For a half-monolayer of carbon monoxide on the copper 
(100) surface, the predicted damping rate of in-plane frustrated translations
is $0.50\times 10^{12}$~s$^{-1}$, as compared to the experimental value
of $(0.43\pm0.07)\times 10^{12}$ s$^{-1}$.  Furthermore it is shown that,
for all coverages presently accessible to experiment, adsorbate motions 
exhibit collective effects which cannot be treated as stemming from 
isolated oscillators.
\end{abstract} 

\pacs{}

Low-frequency adsorbate vibrations have important consequences in many
fundamental chemical processes at surfaces, including reactivity, surface
diffusion, and desorption.  Understanding the dynamical nature of the 
low-frequency modes can provide important insights into the physical behavior 
of these processes.  When the frequency of an adsorbate mode is so low that 
it lies within the phonon band of the substrate, it can resonantly mix with 
the phonons to form new normal modes with composite adsorbate- and bulk-like 
character.  Thus the adsorbate mode can be viewed as developing a finite
lifetime by decaying into propagating bulk elastic waves.  This resonant 
coupling is an important mechanism in understanding energy relaxation processes 
at surfaces.  It has been shown, for example, to dominate the relaxation dynamics 
of the in-plane frustrated translational motion of a carbon monoxide adlayer on 
the (100) surface of copper.~\cite{Lewis96}

In this article, we present an analytical theory of the resonant coupling of
an adsorbate overlayer to its substrate.  This theory results in simple
mathematical expressions for the vibrational damping rate and the 
renormalized vibrational frequency of the adlayer, in terms of fundamental 
properties of the system, such as the mass density and speed of sound of 
the substrate, the mass of the adsorbate molecules, and the adsorbate 
vibrational frequency.  When applied to systems for which there are 
experimental data, these expressions give quantitatively accurate results, 
thus verifying their predictive power.

In this theory, we model the substrate as a flat, isotropic, semi-infinite, 
classical elastic continuum.  The elastic-continuum approximation is valid when 
considering adsorbate modes lying in the frequency range of long-wavelength
substrate acoustic phonons, where the dispersion can be accurately modeled with 
a Debye spectrum.  Furthermore, the choice of a flat and isotropic substrate is 
particularly appropriate for low-index surfaces of close-packed materials.  Thus, 
this theory is especially applicable to the in-plane frustrated-translational 
modes of molecules on low-index metal surfaces.~\cite{Lewis96}\  These modes 
typically have frequencies in the acoustic range of the substrate.  The 
molecular overlayer is modeled as a two-dimensional periodic lattice (or net) 
of identical, classical point masses, each connected to the surface of the 
substrate by a classical harmonic spring.

The theory presented here builds on the earlier work of Persson and Ryberg (PR),
\cite{Persson85}\ who examined resonant vibrational decay for the case of a 
single, isolated adsorbate coupled to an elastic substrate.  They deduced 
the following expression for the damping rate:
\begin{equation}
\gamma = {1\over 8\pi} {m\over \rho} \left({\omega_0 \over c_T} \right)^3 
\omega_0 \xi \;,
\end{equation}
where $\omega_0$\ is the resonance frequency, $m$ is the adsorbate mass, $\rho$ 
is the substrate mass density, $c_T$ is the transverse speed of sound in the 
substrate, and $\xi$ is a dimensionless constant of order unity which depends
only on the ratio of the transverse and longitudinal speeds of sound.  We show 
below, however, that a dense layer of adsorbates behaves fundamentally differently.  
Collective motion of the adlayer strongly enhances the coupling efficiency compared 
to the isolated case, whereas interference effects restrict the phase space of 
substrate modes available for coupling.  One might imagine that both limits could 
be achieved experimentally by varying the surface coverage.  However, we also show 
below that the dense-adlayer limit holds for an enormous range of coverages.  The 
isolated-adsorbate limit does not set in until coverages well below that which
can be observed by current experimental techniques.

We begin by developing the theory in its most general form, which includes both
the dense-adlayer and the isolated-adsorbate extremes as limiting cases.  
Afterwards, we indicate how to deduce the PR result and then focus in detail on 
the dense-adlayer limit, as this is the experimentally realizable case.  In a
forthcoming paper, we will address the general result in greater detail, 
examining the cross-over behavior between the two extremes.


Consider a semi-infinite, classical elastic medium described by the displacement 
field $\bbox{u}(\bbox{x},t)$ and occupying the half-space $z>0$.  Its surface is the 
$xy$-plane.  (In this paper three-dimensional vectors are denoted in boldface,
whereas two-dimensional vectors on the surface are designated with arrows.)  
The surface is covered by a two-dimensional periodic array of point masses $m$ 
attached at lattice sites $\{\vec{R}_i\}$ by classical harmonic springs of bare 
frequency $\omega_0$\ (i.e., spring constant $m\omega_0^2$).  Let $s_i(t)$ denote the 
displacement coordinate of the $i$th adsorbate from its equilibrium position.  
In general, the polarization of $s_i$ can have both in-plane and normal components.  
However, for simplicity, we consider $s_i$ to be fully $x$-polarized.  It can be 
shown, in fact, that the mathematical formulas describing the dynamical response 
of in-plane and normal motion are identical up to numerical constants of order 
unity.  Furthermore, $x$- and $y$-polarized motion are degenerate by symmetry.

The equation of motion for the $i$th adsorbate is given (in the frequency domain)
by:
\begin{equation}
-m\omega^2 s_i(\omega) + m\omega_0^2\left[ s_i(\omega) - u_x(\vec{R}_i,\omega) \right] 
= f_i(\omega)\;,
\label{adseom}
\end{equation}
where $f_i(\omega)$ is the $x$-component of an external driving force on the $i$th 
adsorbate.  If the substrate were rigid (i.e., $\bbox{u}(\bbox{x},\omega) \equiv 0$), 
the adsorbates would clearly undergo simple harmonic motion at frequency $\omega_0$.  
However an {\it elastic} substrate couples to the adsorbate motion, thereby 
damping the vibration and renormalizing its frequency.  The influence of the
coupling on the adsorbate dynamics can be described mathematically by the 
complex poles of the adsorbate response function, $\chi(\vec{R}_i,\omega)$, 
defined by
\begin{equation}
s_i(\omega) = \chi(\vec{R}_i,\omega) f_i(\omega)\;.
\label{defxi}
\end{equation}
To determine $\chi(\vec{R}_i,\omega)$, we first integrate out the substrate degrees 
of freedom, $u_x(\vec{x},\omega)$.

The equation of motion for an elastic substrate subject to an external driving
force, $\bbox{a}(\vec{x},\omega)$, at the surface is~\cite{Landau86}
\begin{equation}
-\omega^2 \bbox{u}(\bbox{x},\omega) - c_L^2 
\nabla\left(\nabla\cdot \bbox{u}(\bbox{x},\omega) \right) 
+ c_T^2 \nabla \times \left( \nabla \times \bbox{u}(\bbox{x},\omega) \right) 
= \delta(z) \bbox{a}(\vec{x},\omega)/\rho\;,
\label{substreom}
\end{equation}
where $c_L$ and $c_T$ are the longitudinal and transverse speeds of sound,
respectively, specifying the elastic properties of the material.
The surface driving force is provided by the oscillating adsorbates, and thus 
is given by
\begin{eqnarray}
a_x(\vec{x},\omega) &=& \sum_i m \omega_0^2 
\left[ s_i(\omega) - u_x(\vec{R}_i,\omega) \right]\,\delta(\vec{x}-\vec{R}_i) 
\label{surftens} \\
a_{y,z}(\vec{x},\omega) &=& 0\;. \nonumber
\end{eqnarray}

The solution to Eqs.~(\ref{substreom}) and (\ref{surftens}) is given formally by
\begin{equation}
u_\alpha(\vec{x},z;\omega) = \sum_\beta 
\int D_{\alpha\beta}(\vec{x},z;\vec{x}',z';\omega)\, a_\beta(\vec{x}',z';\omega)\, 
d^2x' dz'\;,
\label{Green1}
\end{equation}
where $\alpha$ and $\beta$ run over Cartesian directions, and 
$D_{\alpha\beta}(\vec{x},z;\vec{x'},z';\omega)$ is the Green's function of
the elastic substrate, defined as the $\alpha$-component of the solution 
to Eq.~(\ref{substreom}) with source term 
$\delta^2(\vec{x}-\vec{x}')\delta(z-z')/\rho$ in the $\beta$-direction
and stress-free boundary conditions.  
Several simplifications can be made for the present problem.  First, the 
driving force, $\bbox{a}$, exists only at the surface and is $x$-polarized.
Furthermore, according to Eq.~(\ref{adseom}), only the $x$-component of 
$\bbox{u}(\bbox{x},\omega)$ at the surface is of interest.  Finally, translational
symmetry parallel to the surface means that $D_{\alpha\beta}$ can only
depend on the difference $\vec{x}-\vec{x}'$.  Thus, Eq.~(\ref{Green1})
becomes
\begin{equation}
u_x(\vec{x},\omega) = \int D_{xx}(\vec{x}-\vec{x}',\omega)\, a_x(\vec{x}',\omega) 
d^2x'\;,
\label{Green2}
\end{equation}
where the $z$ and $z'$ arguments are zero and have been suppressed for 
convenience.  In reciprocal space,
\begin{equation}
u_x(\vec{k},\omega) = D_{xx}(\vec{k},\omega)\, a_x(\vec{k},\omega)\;.
\label{Green3}
\end{equation}

Because of the two-dimensional periodicity of the adsorbate lattice, the
stationary solutions to both the substrate and adsorbate equations of
motion must obey Bloch's Theorem.  Namely, 
\begin{eqnarray}
u_x(\vec{R}_i,\omega) &=& u_x(0,\omega)\, e^{i \vec{q} \cdot \vec{R}_i} 
\label{Bloch} \\
\nonumber
s_i(\omega) &=& s_0(\omega)\, e^{i \vec{q} \cdot \vec{R}_i}\;,
\end{eqnarray}
where $\vec{q}$ is a wavevector in the first Brillouin zone of the
adsorbate lattice.  Combining Eqs.~(\ref{surftens}) and (\ref{Bloch}) 
and Fourier transforming gives
\begin{eqnarray}
\nonumber
a_x(\vec{k},\omega) &=& m\omega_0^2\, \left[ s_0(\omega) - u_x(0,\omega) \right]
\, \sum_i e^{i (\vec{q}-\vec{k})\cdot \vec{R}_i} \\
&=& m\omega_0^2\, \left[ s_0(\omega) - u_x(0,\omega) \right] \,
{(2\pi)^2 \over A_c} \sum_{\vec{G}} \delta(\vec{q}-\vec{k}+\vec{G})\;,
\label{axk}
\end{eqnarray}
where $A_c$ is the area of the surface unit cell, and $\{\vec{G}\}$ is 
the set of reciprocal lattice vectors of the overlayer structure.

Equation~(\ref{axk}) demonstrates that a coherent excitation of an adsorbate 
overlayer cannot couple to all resonant substrate modes, in contrast to the 
case of an isolated adsorbate.~\cite{Persson85}\  Rather, the lattice imposes 
a selection rule.  Interference among the adsorbates in the overlayer 
prevents energy from being radiated into any substrate mode whose 
in-plane wavevector component is not commensurate with the adsorbate 
reciprocal lattice.  Furthermore, the factor of $1/A_c$ in Eq.~(\ref{axk}) 
shows that dense adlayers drive the substrate more strongly than diffuse 
adlayers, or, in the extreme limit, than an isolated adsorbate.  Thus the 
coupling is more efficient for a dense overlayer and leads to a faster 
damping, as shown below.

Combining Eqs.~(\ref{Green3}) and (\ref{axk}), inverse Fourier transforming, 
and evaluating the result at $\vec{x}=\vec{R}_i$ gives a simple expression 
directly relating the substrate and adsorbate degrees of freedom:
\begin{equation}
u_x(\vec{R}_i,\omega) = \left[ s_i(\omega) - u_x(\vec{R}_i,\omega) \right] \,
{m\omega_0^2 \over A_c} \sum_{\vec{G}} D_{xx}(\vec{q}+\vec{G},\omega) \;.
\label{ux_solv}
\end{equation}
The response function, $\chi(\vec{q},\omega)$, defined as the reciprocal-space
version of the function appearing in Eq.~(\ref{defxi}), can be expressed in terms
of the substrate Green's function, $D_{xx}(\vec{q},\omega)$, and fundamental
parameters of the system by combining Eqs.~(\ref{adseom}) and (\ref{ux_solv}), and
rearranging the result to be in the form of Eq.~(\ref{defxi}).  Thus,
\begin{eqnarray}
\chi(\vec{q},\omega) &=& \left[ -m\omega^2 + { m\omega_0^2 \over 1 + T(\vec{q},\omega) } 
\right]^{-1}   \label{xi_inv} \\
\nonumber
               &=& \left[ -m\omega^2 + m\omega_0^2 \, { 1+R(\vec{q},\omega) \over 
\left( 1+R(\vec{q},\omega) \right)^2 + I^2(\vec{q},\omega)} -i\omega_0 \,
{ m\omega_0 I(\vec{q},\omega) \over \left( 1+R(\vec{q},\omega) \right)^2 + 
I^2(\vec{q},\omega)} \right]^{-1}  \;,
\end{eqnarray}
where
\begin{equation}
T(\vec{q},\omega) = {m\omega_0^2 \over A_c} \sum_{\vec{G}} D_{xx}(\vec{q}+\vec{G},\omega)\;,
\label{Tdef}
\end{equation}
and $R(\vec{q},\omega)$ and $I(\vec{q},\omega)$ are the real and imaginary parts,
respectively, of $T(\vec{q},\omega)$.
The isolated-adsorbate extreme is obtained from Eq.~(\ref{Tdef}) in the limit 
of infinitely long overlayer lattice vectors.  The Brillouin zone area, 
$(2\pi)^2/A_c$, becomes an infinitesimal and the sum over reciprocal lattice 
vectors becomes an integral:~\cite{cfPR}
\begin{equation}
T(\omega) = m\omega_0^2 \int {d^2q \over (2\pi)^2} D_{xx}(\vec{q},\omega)\;.
\label{T_isol}
\end{equation}

The imaginary part of $\chi(\vec{q},\omega)$, ${\rm Im}[\chi(\vec{q},\omega)]$, is the 
measurable absorption spectrum.  If we assume that $T(\vec{q},\omega)$ is sufficiently 
smooth and that the damping rate, $\gamma$, is much smaller than the resonant 
frequency, then ${\rm Im}[\chi(\vec{q},\omega)]$ is a sharply peaked function centered 
at the renormalized frequency $\bar{\omega}_0$ (the experimentally measured frequency) 
and having characteristic width $\gamma$.  
Under these conditions, we can approximate $\omega \approx \omega_0$, which gives 
$\chi(\vec{q},\omega)$ the form of a damped harmonic oscillator:
\begin{equation}
\chi(\vec{q},\omega) \approx \left[ -m\omega^2 + m\bar{\omega}_0^2 -i\omega m\gamma 
\right]^{-1} \;,
\label{dampedHO}
\end{equation}
where
\begin{eqnarray}
\bar{\omega}_0^2 &=& \omega_0^2 \, {1+R(\vec{q},\omega_0) \over \left( 1+R(\vec{q},\omega_0) 
\right)^2 + I^2(\vec{q},\omega_0)} 
\label{wbar0} \\
\gamma &=& { \omega_0 I(\vec{q},\omega_0) \over \left( 1+R(\vec{q},\omega_0) \right)^2 
+ I^2(\vec{q},\omega_0)} \;.
\label{gamma}
\end{eqnarray}
For the remainder of this paper we assume that dispersion in the plane
is negligible.  This assumption is governed by the length scale for 
inter-adsorbate interactions, and is expected to be valid for all but the 
highest surface coverages.  Thus all analyses are made for the $q=0$ limit.
This restriction could be lifted without difficulty, if warranted.

Determining the dynamical response of the adlayer has been reduced to solving
for the Green's function of a flat, semi-infinite, isotropic elastic medium.  
This has been carried out in great detail by Maradudin, {\it et al}.\cite{Maradudin}\
Evaluating their expressions for $D_{xx}$ at $z=z'=0$ gives
\begin{equation}
D_{xx}(\vec{k},\omega) = -i \left[ {k_y^2 \over k^2 \beta p_T} + {k_x^2 \omega^2
  \over 2 k^2 c_T^2}\, {p_T\over p_L} \, {1\over Q(k^2,\omega)} \right]\;,
\label{Dxx}
\end{equation}
where
\begin{eqnarray}
p_{L,T} &=& \left[ {\omega^2 \over c_{L,T}^2} - k^2 \right]^{\frac{1}{2}} \;, 
\label{pLTdef} \\
Q(k^2,\omega) &=& 2\beta k^2 p_T + {(p_T^2 - k^2) \over 2 p_L}
    \left[ (\alpha-\beta)k^2 + (\alpha+\beta)p_L^2 \right] \;,
\label{Qdef}
\end{eqnarray}
and $\alpha=\rho (c_L^2 - c_T^2)$ and $\beta=\rho c_T^2$.  
Equations~(\ref{Dxx})--(\ref{Qdef}) use the same naming conventions 
as in Appendix B of PR.~\cite{Persson85}\  The analysis that has been 
presented so far for $x$-polarized adsorbate motion also holds for 
$z$-polarized motion, provided that $u_z$ and $D_{zz}$ replace $u_x$ 
and $D_{xx}$, respectively, throughout, where $D_{zz}$ is given by
\begin{equation}
D_{zz}(\vec{k},\omega) = -i { k^2 + p_T^2 \over 2Q(k^2,\omega) }\;.
\label{Dzz}
\end{equation}

According to Eq.~(\ref{Tdef}), $D_{xx}(\vec{k},\omega)$ is to be evaluated only 
at reciprocal lattice vectors, $\vec{k}=\vec{G}$.  This function is pure
imaginary unless $p_T$ or $p_L$ is imaginary.
Consider the shortest non-zero reciprocal lattice vector, $\vec{G}_0$.  
Assuming a square adsorbate lattice, $|\vec{G}_0|=2\pi/a$, where $a$ is the
adsorbate lattice constant.  The condition for $p_{T,L}$ to be imaginary is
$\omega/c_{T,L} < |\vec{G}|$.  At $\vec{G}_0$, this is equivalent to
\begin{equation}
a < \lambda_{T,L}\;, 
\label{condition}
\end{equation}
where $\lambda_{T,L} = 2\pi c_{T,L}/\omega$ is the wavelength of a substrate 
phonon at or near the resonant frequency.  However, since the elastic-continuum
description of the substrate is only valid at the long-wavelength end of the
phonon spectrum, Eq.~(\ref{condition}) will be true for the present theory
unless the adsorbate lattice is extremely dilute.  Suppose, for example, that 
$\lambda = 10 a_{\rm bulk}$, where $a_{\rm bulk}$ is the lattice constant of 
the bulk substrate material.  Since surface coverage 
is given by $\theta=a_{\rm bulk}^2/a^2$, Eq.~(\ref{condition}) will be violated 
only for coverages less than 1\%, which is below the resolution limit of current 
surface-sensitive measurement techniques.  Thus, for all systems presently
accessible to experiment, all non-zero reciprocal lattice vectors contribute only
to the real part of $T(\omega)$, and only $\vec{G}=0$ contributes to the imaginary
(i.e., damping) part.  We call the regime specified by Eq.~(\ref{condition})
the ``dense adlayer'' limit, and it is given in terms of surface coverage by
\begin{equation}
\theta > \left( {\omega \over 2 \omega_D} \right)^2 \;,
\label{dense}
\end{equation}
where $\omega_D = c_{T,L}\,\pi / a_{\rm bulk}$ is the substrate Debye frequency.

In the dense adlayer limit, the imaginary part of $T(\omega_0)$ is
\begin{equation}
I(\omega_0) = {m\omega_0^2 \over A_c} D_{xx}(0,\omega_0) = {m\omega_0 \over A_c \rho c_T} \;,
\label{ImT}
\end{equation}
which, when inserted in Eq.~(\ref{gamma}), gives a damping rate of
\begin{equation}
\gamma = {m\omega_0^2 \over A_c \rho c_T}\,
{1 \over \left( 1+R(\omega_0) \right)^2 + I^2(\omega_0)}
= {m\bar{\omega}_0^2 \over A_c \rho c_T}\, \left[{1 \over 1 + R(\omega_0)} \right]
\approx {m\bar{\omega}_0^2 \over A_c \rho c_T}\;,
\label{gamma2}
\end{equation}
where the expression for $\bar{\omega}_0$ in Eq.~(\ref{wbar0}) has been used.  
Typically, $|R(\omega_0)|<<1$, so that the bracketed term in Eq.~(\ref{gamma2}) 
can be neglected.  Thus the vibrational damping rate of a dense adlayer
can be simply computed from empirically available information about the 
system.  Note that the expression for $\gamma$ in Eq.~(\ref{gamma2}) 
is strongly dependent on the surface coverage through the parameters $A_c$,
$\bar{\omega}_0$, and $R(\omega_0)$.

The real part of $T(\omega_0)$ is a sum over all non-zero reciprocal lattice
vectors.  For real systems, the substrate is not a continuum, but rather
a discrete lattice.  This places an upper bound on the summation limit
for the computation of $R(\omega_0)$: the sum includes all reciprocal lattice
vectors of the overlayer structure that lie within the first Brillouin
zone of the {\it bare} surface.  Surface wavevectors outside this region
do not correspond to new substrate phonons.~\cite{Kittel}

The Green's function $D_{xx}(\vec{G},\omega_0)$, for $\vec{G}\neq 0$, can be
simplified in the dense-adlayer limit by expanding to lowest order in 
$(\omega_0/c_{L,T}\, G)^2$, where $G=|\vec{G}|$:
\begin{equation}
D_{xx}(\vec{G},\omega_0) \approx {1 \over \rho c_T^2 G} \, \left[ 
{G_y^2 \over G^2} + {G_x^2 \over G^2}{1\over 2 \left(1-(c_T/c_L)^2 \right)}
\right] \;.
\label{DxxGish}
\end{equation}
Since the term in brackets is of order unity, $R(\omega_0)$ is approximately
given by
\begin{equation}
R(\omega_0) \approx \sum_{\vec{G}\neq 0} {m\omega_0^2 \over A_c \rho c_T^2 G} \;.
\label{Rapprox}
\end{equation}
If the overlayer unit cell is small (i.e., high coverage), then the sum
will contain only a few reciprocal lattice vectors before reaching the 
upper bound set by the substrate lattice.  Thus, since $\omega_0/c_T\, G \ll 1$,
$R(\omega_0)$ will be small.  As the unit cell gets larger, more terms are
included in the sum.  However, the factor $1/A_c$ decreases, keeping
$R(\omega_0)$ small.

%
\begin{table*}[t]
\caption{Empirical physical parameters for a half-monolayer of 
carbon monoxide on the copper (100) surface.
}

\begin{tabular}{ld}
Quantity & Value \\
\hline
& \\[-9pt]
$\bar{\omega}_0$ (10$^{12}$ s$^{-1}$) & 6.03 \tablenote[1]{Reference \protect\onlinecite{Germer}} \\
$m$ (10$^{-23}$ g)                & 4.65 \\
$a$ (10$^{-8}$ cm)                & 3.61 \tablenote[2]{Reference \protect\onlinecite{Kittel}} \\
$A_c=a^2$ (10$^{-15}$ cm$^2$)     & 1.30 \\
$\rho$ (g/cm$^3$)                 & 8.93 \tablenotemark[2] \\
$c_T$ (10$^5$ cm/s)               & 2.91 
\tablenote[3]{Derived from elastic constants in Reference~\protect\onlinecite{Kittel}} \\
$c_L$ (10$^5$ cm/s)               & 4.34 \tablenotemark[3] \\
\end{tabular}

\label{params}
\end{table*}
%

To test the validity of this theory, we now apply it to a system
for which there are experimental data: carbon monoxide on the (100) 
surface of copper at half-monolayer coverage.  The in-plane frustrated
translational mode has a measured frequency of 32 cm$^{-1}$, well within 
the range of long-wavelength copper phonons.  Recently, the lifetime of
this mode was measured to be $2.3\pm0.4$ ps,\cite{Germer}\ which 
corresponds to a damping rate $(0.43\pm0.07)\times 10^{12}$ s$^{-1}$.
Other parameters for this system are compiled in Table \ref{params}.
Using these data, Eq.~(\ref{gamma2}), with the bracketed term neglected,
predicts the frustrated translational damping rate to be 
$\gamma = 0.50\times 10^{12}$~s$^{-1}$, in excellent agreement with 
the experimental value.

The bracketed term in Eq.~(\ref{gamma2}) produces a small downward 
correction.  To show this we now compute $R(\omega_0)$ using the 
expression in Eq.~(\ref{DxxGish}).  The CO overlayer forms a square 
lattice with $c(2\times 2)$ symmetry relative to the underlying 
copper (100) surface.  Therefore, the only non-zero adsorbate reciprocal 
lattice vectors that lie within the first Brillouin zone of the bare 
surface are $[\pm 2\pi/a,0]$ and $[0,\pm 2\pi/a]$, leading to a sum with 
only four terms.  The resulting expression for $R(\omega_0)$ for this 
system is 
\begin{equation}
R(\omega_0) \approx {m\omega_0^2 \over 2\pi \rho c_T^2 a} \,
\left( 2 + {1\over 1-(c_T/c_L)^2} \right) = 0.0376 \;.
\label{R_COCu}
\end{equation}
Using this value, the corrected damping rate is $0.48\times 10^{12}$~s$^{-1}$.
Furthermore, the estimated frequency renormalization factor is 0.956, which, 
as anticipated, is close to unity.  Note that the value of $R(\omega_0)$ was 
computed using the measured frequency and not the bare frequency.  However, 
since the renormalization is small, this approximation is acceptable.  

In previous work, we performed an atomistic investigation of this system
treating the interactions quantum mechanically.~\cite{Lewis96}\  The
computed frustrated translational frequency and damping rate were
27~cm$^{-1}$ and $0.33\times 10^{12}$~s$^{-1}$, respectively.  Thus 
the present elastic continuum theory, the atomistic theory, and the 
experiments all give a consistent, mutually reinforcing picture of the
low-frequency adsorbate vibrational relaxation dynamics; namely, a layer 
of oscillating adsorbates coupling collectively to substrate phonons at 
the resonant frequency.

While much research on adsorbed surfaces has focused on ordered overlayer
structures, it is also important to consider how the generic case of a 
disordered or defective overlayer affects the damping rate formula of 
Eq.~(\ref{gamma2}).  For a dense adlayer (as defined above), the relaxation 
is governed by atomic motions in which all atoms of a given layer oscillate 
in phase, for both the adlayer and the substrate.  Thus the relevant behavior
is one-dimensional along the surface-normal direction, and the details of
the in-plane structure are not resolved.  As a result, the damping rate 
formula, including its coverage and frequency dependence, is unaffected by
disorder in the dense-adlayer limit.  In this analysis we have assumed that
the adsorbates are uniformly distributed on the surface (i.e., no islands).  
If islands form, then the local adsorbate density is higher than the average 
density, and we expect the resonant damping rate to reflect the former value.  
A more thorough, quantitative treatment of disorder in this problem will be 
presented in a forthcoming article.

Resonant coupling between an adsorbate overlayer and an elastic substrate 
provides a damping force for the decay of low-frequency vibrations.  We 
have presented a classical, analytical theory of this mechanism which 
predicts vibrational damping rates in excellent quantitative agreement 
with measured values for systems that have been studied experimentally.

Furthermore, we have shown that, even for the smallest measurable coverages, 
vibrating adsorbates behave collectively when coupling to the substrate.  Due 
to interference effects, a layer of adsorbates oscillating in phase couples
only to resonant phonon modes propagating perpendicular to the surface.  Only 
when neighboring adsorbates are separated by distances larger than the 
wavelength of resonant phonons will new propagation directions be included.  
The isolated-adsorbate limit, treated by Persson and Ryberg,~\cite{Persson85}\ 
becomes valid only when {\it many} resonant-phonon wavelengths fit between 
neighboring adsorbate sites.  This limit is well below the measurable 
threshold.  Finally, we have also shown that the coupling is strongly 
coverage dependent.  A dense overlayer drives the substrate more efficiently 
than does a dilute overlayer.

We thank J. P. Toennies for useful discussions during the early stages of
this work.  Financial support for this project was provided by the National
Science Foundation under grants DMR 97-02514 and DMR 93-13047, and by the
Petroleum Research Fund of the American Chemical Society under grant 
32007-G5.


\begin{thebibliography}{MM}

\bibitem{Lewis96} S. P. Lewis and A. M. Rappe, Physical Review Letters {\bf 77},
5241 (1996).

\bibitem{Persson85} B. N. J. Persson and R. Ryberg, Physical Review B {\bf 32},
3586 (1985).

\bibitem{Landau86} L. D. Landau and E. M. Lifshitz, {\it Theory of Elasticity,}
3rd ed.\ (Pergamon, New York, 1986).

\bibitem{cfPR} The function $T(\omega)$ in Eq.~(\ref{T_isol}) is the negative of
the function $R(\omega)$ from Appendix B of Ref.~\onlinecite{Persson85}.

\bibitem{Maradudin} A. A. Maradudin and D. L. Mills, Annals of Physics 
{\bf 100}, 262 (1976); and M. G. Cottam and A. A. Maradudin, {\it Surface 
Excitations}, edited by V. M. Agranovich and R. Loudon (Elsevier, New York, 1984).

\bibitem{Kittel} C. Kittel, {\it Introduction to Solid State Physics,} 7th ed.\
(Wiley, New York, 1996).

\bibitem{Germer} T. A. Germer, J. C. Stephenson, E. J. Heilweil, and R. R.
Cavanagh, Physical Review Letters {\bf 71}, 3327 (1993); Journal of Chemical 
Physics {\bf 101}, 1704 (1994).

\end{thebibliography}
\end{document}